\documentstyle[12pt,a4]{article}
\input epsf
\global\arraycolsep=2pt

\begin{document}

\begin{titlepage}

\begin{flushright}
CERN-TH/95-107\\
hep-ph/9505238\\
May 1995
\end{flushright}

\vspace{0.5cm}

\begin{center}
\Large\bf Uncertainties in the Determination of $|V_{cb}|$
\end{center}

\vspace{1.0cm}

\begin{center}
Matthias Neubert\\
{\sl Theory Division, CERN, CH-1211 Geneva 23, Switzerland}
\end{center}

\vspace{1.2cm}

\begin{abstract}
I discuss the theoretical uncertainties in the extraction of
$|\,V_{cb}|$ from semilep\-to\-nic decays of $B$ mesons, taking into
account the most recent theoretical developments. The main sources of
uncertainty are identified both for the exclusive decay mode $B\to
D^*\ell\,\bar\nu$ and for the inclusive channel $B\to
X\,\ell\,\bar\nu$. From an analysis of the available experimental
data, I obtain $|\,V_{cb}|_{\rm excl} = 0.041\pm 0.003_{\rm exp}\pm
0.002_{\rm th}$ from the exclusive mode, and $|\,V_{cb}|_{\rm incl} =
0.040\pm 0.001_{\rm exp}\pm 0.005_{\rm th}$ from the inclusive mode.
I also give a prediction for the slope of the form factor
${\cal{F}}(w)$ at zero recoil, which is $\widehat\varrho^2=0.8\pm
0.3$.
\end{abstract}

\vspace{1.0cm}

\centerline{\it to appear in the Proceedings of the XXXth Rencontres
de Moriond}
\centerline{\it ``Electroweak Interactions and Unified Theories''}
\centerline{\it Les Arcs, France, March 1995}

\vspace{2.0cm}

\noindent
CERN-TH/95-107\\
May 1995

\end{titlepage}

\section{Introduction}

\begin{figure}[htb]
   \vspace{0.5cm}
   \epsfysize=5cm
   \centerline{\epsffile{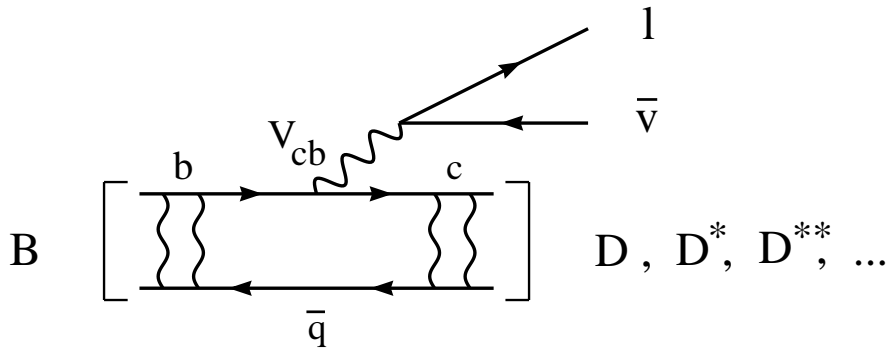}}
   \centerline{\parbox{13cm}{\caption{\label{fig:1}
Semileptonic decay of a $B$ meson.}}}
\end{figure}

Semileptonic decays of $B$ mesons have received a lot of attention in
recent years. The decay channel $B\to D^*\ell\,\bar\nu$ has the
largest branching fraction of all $B$-meson decay modes, and large
data samples have been collected by various experimental groups. From
the theoretical point of view, semileptonic decays are simple enough
to allow for a reliable, quantitative description. Yet, the analysis
of these decays provides much information about the strong forces
that bind the quarks and gluons into hadrons. Schematically, a
semileptonic decay process is shown in Fig.~\ref{fig:1}. The strength
of the $b\to c$ transition vertex is governed by the element $V_{cb}$
of the Cabibbo--Kobayashi--Maskawa (CKM) matrix. The parameters of
this matrix are fundamental parameters of the Standard Model. A
primary goal of the study of semileptonic decays of $B$ mesons is to
extract with high precision the values of $V_{cb}$ and $V_{ub}$. The
problem is that the Standard Model Lagrangian describing these
transitions is formulated in terms of quark and gluon fields, whereas
the physical hadrons are bound states of these degrees of freedom.
Hence, an understanding of the transition from the quark to the
hadron world is necessary before the fundamental parameters can be
extracted from experimental data.

Today, our knowledge of the elements of the CKM matrix, as extracted
from direct measurements of flavour-changing transitions, is as
follows: The best known entries are $V_{ud}$ and $V_{us}$, which have
an uncertainty of $0.1\%$ and 1\%, respectively. The next well known
entry is already related to the $b$-quark; $V_{cb}$ is now determined
to an accuracy of 7\%. Then follow $V_{cd}$, $V_{cs}$, and $V_{ub}$,
with 10\%, 20\%, and 30\% uncertainties, respectively. No direct
measurements exist for the matrix elements related to the top quark.

In this talk I discuss the status of the theoretical developments
underlying the determination of $V_{cb}$, both from exclusive and
from inclusive semileptonic decays of $B$ mesons.

\section{$|\,V_{cb}|$ from Exclusive Decays}

With the discovery of heavy-quark symmetry (for a review see
Ref.~\cite{review} and references therein), it has become clear that
the study of the exclusive semileptonic decay mode $\bar B\to
D^*\ell\,\bar\nu$ allows for a reliable determination of
$|\,V_{cb}|$, which is free, to a large extent, of hadronic
uncertainties \cite{Volo}--\cite{new}. Model dependence enters this
analysis only at the level of power corrections of order
$(\Lambda_{\rm QCD}/m_Q)^2$.\footnote{I shall use $\Lambda_{\rm
QCD}\sim 0.25$~GeV as a characteristic low-energy scale of the strong
interactions, and $m_Q$ as a generic notation for $m_c$ or $m_b$.}
These corrections can be investigated in a systematic way, using the
heavy-quark effective theory \cite{Geor}. They are found to be small,
of order a few per cent.

The analysis consists in measuring the recoil spectrum in the decay
$B\to D^*\ell\,\bar\nu$. One introduces the kinematic variable
\begin{equation}
   w = v_B\cdot v_{D^*} = {E_{D^*}\over m_{D^*}}
   = {m_B^2 + m_{D^*}^2 - q^2\over 2 m_B m_{D^*}} \,,
\end{equation}
which is the product of the four-velocities of the mesons. Here
$E_{D^*}$ denotes the recoil energy of the $D^*$ meson in the parent
rest frame, and $q^2=(p_B-p_{D^*})^2$ is the invariant momentum
transfer. The differential decay rate is given by \cite{Vcb,new}
\begin{eqnarray}\label{BDrate}
   {{\rm d}\Gamma\over{\rm d}w}
   &=& {G_F^2\over 48\pi^3}\,(m_B-m_{D^*})^2\,m_{D^*}^3
    \sqrt{w^2-1}\,(w+1)^2 \nonumber\\
   &&\mbox{}\times \bigg[ 1 + {4w\over w+1}\,
    {m_B^2-2w\,m_B m_{D^*} + m_{D^*}^2\over(m_B - m_{D^*})^2}
    \bigg]\,|\,V_{cb}|^2\,{\cal{F}}^2(w) \,.
\end{eqnarray}
The function ${\cal{F}}(w)$ denotes the (suitably defined) hadronic
form factor for this decay. It is conventional to factorize it in the
form ${\cal{F}}(w)=\eta_A\,\widehat\xi(w)$, where $\eta_A$ is a
short-distance coefficient, and the function $\widehat\xi(w)$
contains the long-distance hadronic dynamics. Apart from corrections
of order $\Lambda_{\rm QCD}/m_Q$, this function coincides with the
Isgur--Wise form factor \cite{Isgu,Falk}. In analogy to the case of
light-quark SU(3) flavour symmetry, in which the Ademollo--Gatto
theorem protects the $K\to\pi$ transition form factor against
first-order symmetry-breaking corrections at $q^2=0$ \cite{AGTh},
there is a theorem which protects the function $\widehat\xi(w)$
against first-order $\Lambda_{\rm QCD}/m_Q$ corrections at the
kinematic point of zero recoil ($w=1$). This is Luke's theorem
\cite{Luke}, which determines the normalization of $\widehat\xi(w)$
at $w=1$ up to corrections of order $(\Lambda_{\rm QCD}/m_Q)^2$,
i.e.\ $\widehat\xi(1)=1+\delta_{1/m^2}$.

The strategy for a precise determination of $|\,V_{cb}|$ is thus to
extract the product $|\,V_{cb}|\,{\cal{F}}(w)$ from a measurement of
the differential decay rate, and to extrapolate it to $w=1$ to
measure
\begin{equation}
   |\,V_{cb}|\,{\cal{F}}(1) = |\,V_{cb}|\,\eta_A\,
   (1 + \delta_{1/m^2}) \,.
\end{equation}
The task of theorists is to provide a reliable calculation of the
quantities $\eta_A$ and $\delta_{1/m^2}$ in order to turn this
measurement into a precise determination of $|\,V_{cb}|$. I will now
discuss the status of these calculations.

\vskip 0.8cm
\noindent
{\large\it 2.1~~~Perturbative corrections}
\vskip 0.3cm

\noindent
The short-distance coefficient $\eta_A$ takes into account the finite
renormalization of the axial vector current arising from virtual
gluon exchange. It can be calculated in perturbation theory. At the
one-loop order, one finds \cite{Pasc,Volo,QCD1}
\begin{equation}
   \eta_A = 1 + {\alpha_s(M)\over\pi}\,\bigg(
   {m_b + m_c\over m_b - m_c}\,\ln{m_b\over m_c} - {8\over 3}
   \bigg) \,.
\end{equation}
The scale $M$ of the running coupling constant is not determined at
this order. A reasonable choice is to take the average virtuality of
the gluon in the one-loop diagrams \cite{BLM}. In the case of
$\eta_A$, this so-called BLM scale has been calculated to be
$M=0.51\sqrt{m_b m_c}$ (in the $\overline{\rm MS}$ scheme)
\cite{etaVA}. Taking then $m_b=4.8$~GeV, $m_c/m_b=0.30\pm 0.05$, and
$\Lambda_{\rm QCD}=0.25$~GeV for the scale parameter in the two-loop
expression for the running coupling constant, one obtains values in
the range $\eta_A=0.950$--0.965.

Several (partial) higher-order calculations have been performed to
improve this result. Using renormalization-group techniques,
logarithms of the type $(\alpha_s\ln z)^n$, $\alpha_s(\alpha_s\ln
z)^n$, and $(m_c/m_b)(\alpha_s\ln z)^n$, where $z=m_c/m_b$, have been
resummed to all orders in perturbation theory
\cite{PoWi}--\cite{QCD2}. This leads to the somewhat larger value
$\eta_A\simeq 0.985$. Another class of higher-order corrections
consists of the so-called renormalon chain contributions, which are
terms of order $\beta_0^{n-1}\alpha_s^n$ in the perturbative series
for $\eta_A$. Resumming these terms to all orders gives the lower
value $\eta_A\simeq 0.945$ \cite{part1}.

The main virtue of these partial higher-order calculations is to
provide an estimate of the theoretical uncertainty in the value of
$\eta_A$. Thus, I quote the final result as
\begin{equation}
   \eta_A = 0.965\pm 0.020 \,.
\end{equation}

\vskip 0.8cm
\noindent
{\large\it 2.2~~~Power corrections}
\vskip 0.3cm

\noindent
Hadronic uncertainties enter the determination of $|\,V_{cb}|$ at the
level of second-order power corrections, which are expected to be of
order $(\Lambda_{\rm QCD}/m_c)^2\sim 3\%$. For a precision
measurement, it is important to understand the structure of these
corrections in detail. This is the most complicated aspect of the
theoretical analysis, which unavoidably introduces some amount of
model dependence. However, since the goal is to estimate an effect
which by itself is very small, even a large relative error in
$\delta_{1/m^2}$ is acceptable.

Three approaches have been suggested to estimate these corrections.
The idea of the ``exclusive'' approach of Falk and myself is to
classify all $1/m_Q^2$ operators in the heavy-quark effective theory
and to estimate their matrix elements between meson states
\cite{FaNe}. This last step is model-dependent. A typical result
obtained in this way is $\delta_{1/m^2}=-(3\pm 2)\%$. In
Ref.~\cite{review}, the error has been increased to $\pm 4\%$ in
order to account for the model dependence and unknown higher-order
corrections. A similar result, $-5\%<\delta_{1/m^2}<0$, has been
obtained by Mannel \cite{Mann}.

The idea of the ``inclusive'' approach of Shifman et al.\ is to apply
the operator product expansion to the $B$-meson matrix element of the
time-ordered product of two flavour-changing currents, and to equate
the resulting theoretical expression to a phenomenological expression
obtained by saturating the matrix element with physical intermediate
states \cite{Shif}. This leads to sum rules that imply inequalities
for the $B\to D^*$ transition form factors. In particular, one
obtains the bound $\delta_{1/m^2}<-\frac{1}{8}\,(m_{B^*}^2-m_B^2)/
m_c^2\simeq -3\%$. The authors of Ref.~\cite{Shif} make an ``educated
guess'' that the value of $\delta_{1/m^2}$ is actually much larger,
namely $-(9\pm 3)\%$.

It is possible to combine the above predictions in a ``hybrid''
approach, which uses the sum rules to put bounds on the hadronic
parameters that enter the ``exclusive'' analysis \cite{new}. One then
finds that for all reasonable choices of parameters the results are
in the range $-8\%<\delta_{1/m^2}<-3\% $, which is consistent with
all previous estimates at the $1\sigma$ level. Thus, I quote the
final result as
\begin{equation}
   \delta_{1/m^2} = -(5.5\pm 2.5)\% \,.
\end{equation}

\vskip 0.8cm
\noindent
{\large\it 2.3~~~Determination of\/ $|\,V_{cb}|$}
\vskip 0.3cm

\noindent
Combining the above results, I obtain for the normalization of the
hadronic form factor at zero recoil
\begin{equation}\label{etaxi}
   {\cal{F}}(1) = \eta_A\,(1 + \delta_{1/m^2})
   = 0.91\pm 0.04 \,.
\end{equation}
To be conservative, I have added the theoretical errors linearly.
Three experiments have recently presented new measurements of the
product $|\,V_{cb}|\,{\cal{F}}(1)$. When rescaled using the new
lifetime values $\tau_{B^0}=(1.61\pm 0.08)$~ps and
$\tau_{B^+}=(1.65\pm 0.07)$~ps \cite{Roud}, the results are
\begin{equation}
   |\,V_{cb}|\,{\cal{F}}(1) = \left\{
   \begin{array}{ll}
   0.0351\pm 0.0019\pm 0.0020 &
    ;\quad \mbox{Ref.~\protect\cite{CLEO},} \\
   0.0364\pm 0.0042\pm 0.0031 &
    ;\quad \mbox{Ref.~\protect\cite{ALEPH},} \\
   0.0388\pm 0.0043\pm 0.0025 &
    ;\quad \mbox{Ref.~\protect\cite{ARGUS},}
   \end{array} \right.
\end{equation}
where the first error is statistical and the second one systematic. I
will follow the suggestion of Ref.~\cite{Ritch} and add $0.001\pm
0.001$ to these values to account for a small positive
curvature\footnote{The fit results are obtained assuming a linear
form of ${\cal{F}}(w)$.}
of the function ${\cal{F}}(w)$. Taking the weighted average of the
experimental results, which is $|\,V_{cb}|\,{\cal{F}}(1)=0.0370\pm
0.0025$, and using the theoretical prediction (\ref{etaxi}), I then
obtain
\begin{equation}
   |\,V_{cb}| = 0.0407\pm 0.0027_{\rm exp}\pm 0.0016_{\rm th} \,,
\end{equation}
which corresponds to a measurement of $|\,V_{cb}|$ with 7\% accuracy.

\vskip 0.8cm
\noindent
{\large\it 2.4~~~Prediction for the slope parameter
$\widehat\varrho^2$}
\vskip 0.3cm

\noindent
In the extrapolation of the differential decay rate (\ref{BDrate}) to
zero recoil, the slope of the function ${\cal{F}}(w)$ close to $w=1$
plays an important role. One defines a parameter $\widehat\varrho^2$
by
\begin{equation}
   {\cal{F}}(w) = {\cal{F}}(1)\,\Big\{ 1
   - \widehat\varrho^2\,(w-1) + \ldots \Big\} \,.
\end{equation}
It is important to distinguish $\widehat\varrho^2$ from the slope
parameter $\varrho^2$ of the Isgur--Wise function. They differ by
corrections that break the heavy-quark symmetry. Whereas the slope of
the Isgur--Wise function is a universal, mass-independent parameter,
the slope of the physical form factor depends on logarithms and
inverse powers of the heavy-quark masses.
The relation between the two parameters is
\cite{new}
\begin{equation}\label{rhorel}
   \widehat\varrho^2 = \varrho^2 + (0.16\pm 0.02) + O(1/m_Q) \,.
\end{equation}
An estimate of the $\Lambda_{\rm QCD}/m_Q$ corrections to this
relation is model-dependent and thus has a large theoretical
uncertainty. I shall not attempt it here.

\begin{figure}[htb]
   \vspace{0.5cm}
   \epsfysize=6cm
   \centerline{\epsffile{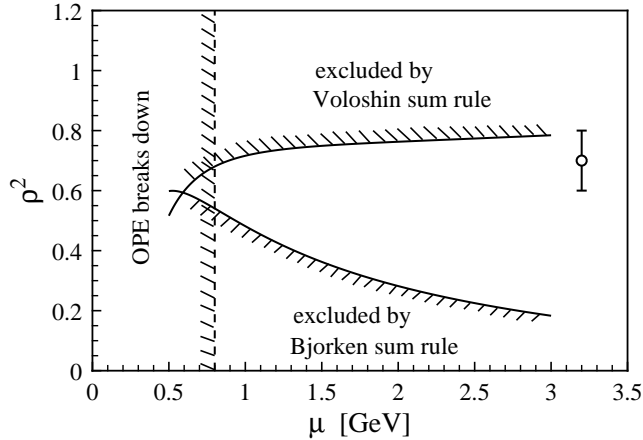}}
   \centerline{\parbox{13cm}{\caption{\label{fig:2}
Bounds for the slope parameter $\varrho^2$ following from the Bjorken
and Voloshin sum rules, from Ref.~\protect\cite{KoNe}. The point with
the error bar shows the QCD sum rule prediction.}}}
\end{figure}

The slope parameter of the Isgur--Wise function, $\varrho^2$, is
constrained by the Bjorken \cite{Bjsum,IsgW} and Voloshin
\cite{Volsum} sum rules. At the tree level, it was known for a long
time that $1/4<\varrho^2<\,\approx 0.75$. However, only recently
Grozin and Korchemsky have shown how to include perturbative and
nonperturbative corrections to these bounds \cite{GrKo,KoNe}. The
results are shown in Fig.~\ref{fig:2}. Here, the scale parameter
$\mu$ has to be chosen large enough for the operator product
expansion to be well defined, but it is otherwise arbitrary. Assuming
that values $\mu>0.8$~GeV are sufficiently large, one finds that
$\varrho^2$ is constrained to be very close to 0.6. This value is in
good agreement with earlier predictions obtained from QCD sum rules,
which gave $\varrho^2=0.7\pm 0.1$ \cite{review,Baga,twoloop}.

{}From Fig.~\ref{fig:2}, and using (\ref{rhorel}), I conclude that
\begin{equation}
   \varrho^2 = 0.65\pm 0.15 \,,\qquad
   \widehat\varrho^2 = 0.8\pm 0.3 \,.
\end{equation}
This prediction compares well with the average value observed
experimentally, which is $\widehat\varrho^2=0.87\pm 0.12$
\cite{CLEO}--\cite{ARGUS}.

\section{$|\,V_{cb}|$ from Inclusive Decays}

Complementary to the analysis of exclusive decays is the study of the
inclusive semileptonic decay rate for $B\to X\,\ell\,\bar\nu$. Since
$|\,V_{ub}/V_{cb}|^2<1\%$, one can to very good approximation neglect
the contribution of charmless final states and consider $X$ to be a
hadronic state containing a charm particle. An obvious advantage of
inclusive decays is the existence of high-statistics data samples.
{}From the theoretical point of view, summing over many final states
eliminates part of the hadronic uncertainty.

As in the exclusive case, the framework for the theoretical
description of inclusive decays is provided by the heavy-quark
expansion. It could be shown that the leading term in this expansion
reproduces the free-quark decay model, while the nonperturbative
corrections to this model can be systematically included in an
expansion in powers of $1/m_b$ \cite{Chay}--\cite{shape}. The total
semileptonic decay rate can be written as
\begin{eqnarray}\label{Gamtot}
   \Gamma = {G_F^2\over 192\pi^3}\,|\,V_{cb}|^2\,m_b^5\,&\Bigg\{&
    \bigg( 1 + {\lambda_1 + 3\lambda_2\over 2 m_b^2} \bigg)\,
    f(m_c/m_b) - {6\lambda_2\over m_b^2}\,\bigg(
    1 - {m_c^2\over m_b^2} \bigg)^4 \nonumber\\
   &&\mbox{}+ {\alpha_s(M)\over\pi}\,g(m_c/m_b) + \dots \Bigg\} \,,
\end{eqnarray}
where the ellipsis represents terms of higher order in $1/m_b$ or
$\alpha_s$. In this expression, $m_b$ and $m_c$ denote the pole
masses
(defined to the appropriate order in perturbation theory) of the
heavy quarks, $f(m_c/m_b)\simeq 0.52$ and $g(m_c/m_b)\simeq -0.87$
are kinematic functions, and $\lambda_1$ and $\lambda_2$ are
nonperturbative hadronic parameters. I will now discuss the
theoretical uncertainties in the evaluation of (\ref{Gamtot}).

\vskip 0.8cm
\noindent
{\large\it 3.1~~~Perturbative corrections}
\vskip 0.3cm

\noindent
Let me first discuss the uncertainty due to unknown higher-order
perturbative corrections. Only the correction of order $\alpha_s$ is
known exactly \cite{Nir}. However, recently Luke et al.\ have
computed the part of the next-order term that depends on the number
of light-quark flavours \cite{LSW}. In the $\overline{\rm MS}$
scheme, the result is
\begin{equation}\label{Gamser}
   {\Gamma\over\Gamma_0} = 1 - 1.67\,{\alpha_s(m_b)\over\pi}
   - (1.68\,\beta_0+\dots)\,\bigg( {\alpha_s(m_b)\over\pi}
   \bigg)^2 + \dots = 1 - 0.11 - 0.06 - \dots \,,
\end{equation}
where $\Gamma_0$ is the decay rate at the tree level, and
$\beta_0=11-\frac{2}{3}\,n_f$ is the first coefficient of the
$\beta$-function. If one uses this partial calculation to estimate
the uncertainty, which for an asymptotic series is given by the size
of the last term to be included, one finds
$(\delta\Gamma/\Gamma)_{\rm pert}\simeq 6\%$, which is in good
agreement with an estimate of the renormalization-scale and -scheme
dependence by Ball and Nierste \cite{BaNi}. Recently, Ball et al.\
have performed an all-order resummation of the terms of order
$\beta_0^{n-1}\alpha_s^n$ for the above series \cite{BBB}. They find
that the effect of higher-order terms is important and leads to
$\Gamma/\Gamma_0=0.77\pm 0.05$. Note that this corresponds to an
effective scale $M\sim 1$~GeV in (\ref{Gamtot}), which is rather low.
The corresponding value of the coupling constant is
$\alpha_s(M)\simeq 0.43$. It is difficult to derive a reliable error
estimate from these analyses, but I think a reasonable number is
\begin{equation}
   \bigg( {\delta\Gamma\over\Gamma} \bigg)_{\rm pert}
   \simeq 10\% \,.
\end{equation}

\vskip 0.8cm
\noindent
{\large\it 3.2~~~Power corrections}
\vskip 0.3cm

\noindent
The leading power corrections in the expression for the inclusive
decay rate appear at order $1/m_b^2$. They are proportional to two
hadronic parameters with a simple physical interpretation:
$\lambda_1$ is related to the average momentum of the $b$-quark
inside a $B$ meson at rest, and $\lambda_2$ is proportional to the
vector--pseudoscalar mass splitting. I shall use
\begin{eqnarray}
   \lambda_1 &=& -\langle\,\vec p_b^{\,2}\rangle
    = -(0.4\pm 0.2)~\mbox{GeV}^2 \,, \nonumber\\
   \lambda_2 &=& {m_{B^*}^2 - m_B^2\over 4}
    = 0.12~\mbox{GeV}^2 \,.
\end{eqnarray}
The value of $\lambda_1$ is a compromise between the theoretical
estimates obtained in Refs.~\cite{BaBr,virial}.

The power corrections reduce the total decay rate by $-(4.2\pm
0.5)\%$, which is a rather small effect. The uncertainty in the value
of $\lambda_1$ introduces an uncertainty on the value of $\Gamma$ of
0.6\%, which is almost negligible. I increase this value in order to
account for higher-order power corrections, which I expect to be of
order $(\Lambda_{\rm QCD}/m_c)^3\sim 0.5\%$, and quote
\begin{equation}
   \bigg( {\delta\Gamma\over\Gamma} \bigg)_{\rm power}
   \simeq 1\% \,.
\end{equation}
This is the smallest contribution to the total theoretical
uncertainty.

\vskip 0.8cm
\noindent
{\large\it 3.3~~~Dependence on quark masses}
\vskip 0.3cm

\noindent
Another source of nonperturbative uncertainty results from the
appearance of the heavy-quark masses in the expression for the
inclusive decay rate. The pole masses of the bottom and charm quarks
have an uncertainty of at least several hundred MeV. Since the rate
is proportional to $m_b^5$, this seems to be a severe limitation.
However, it has been pointed out that the actual uncertainty is lower
due to a strong correlation between the values of the two heavy-quark
masses \cite{Shif}. In fact, one should consider the decay rate as a
function of $m_b$ and of the difference $\Delta m=m_b-m_c$. I shall
assume that $m_b$ has an uncertainty of 300~MeV. However, the mass
difference is known to much higher precision. Using the heavy-quark
expansion, one can derive that \cite{FaNe}
\begin{equation}
   \Delta m = m_b-m_c = (\overline{m}_B-\overline{m}_D)\,\bigg\{
   1 - {\lambda_1\over 2\overline{m}_B\overline{m}_D}
   + O(1/m_Q^3) \bigg\} \,,
\end{equation}
where $\overline{m}_B=\frac{1}{4}\,(m_B+3 m_{B^*})=5.31$~GeV and
$\overline{m}_D=\frac{1}{4}\,(m_D+3 m_{D^*})=1.97$~GeV denote the
spin-averaged meson masses. This relation leads to
\begin{equation}
   \Delta m = (3.40\pm 0.03\pm 0.03)~\mbox{GeV} \,,
\end{equation}
where the first error reflects the uncertainty in the value of
$\lambda_1$, and the second one takes into account unknown
higher-order corrections. Hereafter, I shall assume an uncertainty of
60~MeV in the value of $\Delta m$.

\begin{figure}[htb]
   \vspace{0.5cm}
   \epsfysize=7cm
   \centerline{\epsffile{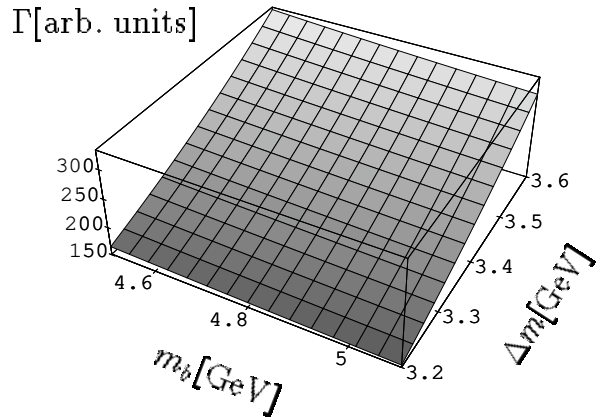}}
   \centerline{\parbox{13cm}{\caption{\label{fig:3}
Dependence of the inclusive semileptonic decay rate $\Gamma$ on the
parameters $m_b$ and $\Delta m=m_b-m_c$.}}}
\end{figure}

In Fig.~\ref{fig:3}, I show the dependence of the decay rate on these
two parameters, using the value $\alpha_s(M)=0.4$ for the strong
coupling constant in (\ref{Gamtot}). Clearly, the variation with
$\Delta m$ is much stronger than the variation with $m_b$. For
$m_b=4.8$~GeV and $\Delta m=3.4$~GeV, I find the partial derivatives
$\delta\Gamma/\Gamma\simeq 5.73\,\delta(\Delta m)/\Delta m$ and
$\delta\Gamma/\Gamma\simeq -0.55\,\delta m_b/m_b$. Since the errors
in $m_b$ and $\Delta m$ are essentially uncorrelated, this leads to
\begin{equation}
   \bigg( {\delta\Gamma\over\Gamma} \bigg)_{\rm masses}
   = \sqrt{ \bigg( 0.101\,{\delta(\Delta m)\over 60~\mbox{MeV}}
   \bigg)^2 + \bigg( 0.034\,{\delta m_b\over 300~\mbox{MeV}}
   \bigg)^2 } \simeq 11\% \,.
\end{equation}
Note that this is dominated by the (rather small) uncertainty in the
mass difference $\Delta m$.

\vskip 0.8cm
\noindent
{\large\it 3.4~~~Determination of\/ $|\,V_{cb}|$}
\vskip 0.3cm

\noindent
Adding the above errors linearly and taking the square root, I
conclude that the theoretical uncertainty in the extraction of
$|\,V_{cb}|$ from inclusive decays is
\begin{equation}
   {\delta|\,V_{cb}|\over|\,V_{cb}|} \simeq 11\% \,.
\end{equation}
{}From an analysis of the experimental data, one then obtains
\cite{Ritch}
\begin{equation}
   |\,V_{cb}| = \left\{
   \begin{array}{ll}
   0.039\pm 0.001_{\rm exp}\pm 0.005_{\rm th} &
    ;\quad \mbox{measurements at $\Upsilon(4s)$,} \\
   0.042\pm 0.002_{\rm exp}\pm 0.005_{\rm th} &
    ;\quad \mbox{measurements at $Z^0$.}
   \end{array} \right.
\end{equation}
The theoretical uncertainty in these numbers is somewhat larger than
in the case of the exclusive analysis; however, the experimental
errors are smaller.

\section{Summary}

The most precise measurements of the element $V_{cb}$ of the CKM
matrix come from the analysis of semileptonic decays of $B$ mesons.
{}From the measurement of the recoil spectrum in the exclusive
channel $B\to D^*\ell\,\bar\nu$, one obtains
\begin{equation}
   |\,V_{cb}|_{\rm excl} = 0.041\pm 0.003_{\rm exp}
   \pm 0.002_{\rm th} \,,
\end{equation}
where the theoretical error is dominated by the uncertainty in the
calculation of nonperturbative power corrections of order
$(\Lambda_{\rm QCD}/m_Q)^2$. From the measurement of the total
inclusive decay rate, on the other hand, one finds
\begin{equation}
   |\,V_{cb}|_{\rm incl} = 0.040\pm 0.001_{\rm exp}
   \pm 0.005_{\rm th} \,.
\end{equation}
In this case, the main theoretical uncertainty comes from the
uncertainty in the value of the quark mass difference $m_b-m_c$, as
well as from higher-order perturbative corrections.

Given the fact that both methods are very different both from the
experimental and from the theoretical point of view, it is most
satisfying that the results are in perfect agreement. Combining them,
I obtain the final value
\begin{equation}
   |\,V_{cb}| = 0.041\pm 0.003 \,.
\end{equation}

\end{document}